\documentclass[12pt]{article}
\usepackage{latexsym}
\usepackage{amsmath}
 \usepackage[english]{babel}
\renewcommand{\epsilon}{\varepsilon}

\usepackage{amsmath,bm}
\usepackage{amssymb}
\usepackage{xcolor}
\usepackage{amsthm}
\usepackage{mathtools}
\usepackage[ansinew]{inputenc}
\usepackage{graphicx}
\usepackage{epsfig}
\usepackage{dsfont}
\usepackage{bbm}
\usepackage{url}
\usepackage{placeins}
\usepackage[FIGTOPCAP]{subfigure} 
\usepackage{multirow}
\usepackage[font=footnotesize,labelfont=bf]{caption}
\usepackage{authblk}
\usepackage{longtable}

\usepackage[square,numbers]{natbib}
\def\3{\ss}

\newcommand{\bea}{\begin{eqnarray*}}
	\newcommand{\eea}{\end{eqnarray*}}
\newcommand{\be}{\begin{eqnarray}}
	\newcommand{\ee}{\end{eqnarray}}

\newcommand{\var}{ \mbox{\sl Var} \ }

\newcommand{\ba}{\begin{array}}
	\newcommand{\ea}{\end{array}}

\newcommand{\Cov}{\text{\rm Cov}}

\def\3{\ss}

\begin{document}
	
	\title{{\bf \Large A non-parametric proportional risk model to assess a treatment effect in time-to-event data
   }}

	\author[1]{L. Ameis}
	\author[2]{O. Kuss} 
	\author[3]{A. Hoyer}  
	\author[1*]{K. M\"ollenhoff}

	\small\affil[1]{\small Mathematical Institute, Heinrich Heine University D\"usseldorf, D\"usseldorf, Germany}
	\affil[2]{\small German Diabetes Center, Leibniz Institute for Diabetes Research at Heinrich Heine University D\"usseldorf, Institute for Biometrics and Epidemiology,	D\"usseldorf, Germany}
	\affil[3]{\small Biostatistics and Medical Biometry, Medical School OWL, Bielefeld University, Bielefeld, Germany}\normalsize

	\date{}
	\pdfminorversion=4
	\maketitle
	
		\begin{abstract}
		Time-to-event analysis often relies on prior parametric assumptions, or, if a non-parametric approach is chosen, Cox's model. This is inherently tied to the assumption of proportional hazards, with the analysis potentially invalidated if this assumption is not fulfilled.  
  %  which is inherently tied to the assumption of proportional hazards. In case of any violation, the quality of the analysis can be reduced significantly \textcolor{red}{OK: Mit diesem Satz kann ich wenig anfangen. Was bedeutet "quality" einer Analyse? Und was bedeutet eine signifikante Reduktion? Ich nehem an, hier ist nicht "stat. signifikant" gemeint ...}. Particularly the assumption of proportional hazards was recently criticized for being rarely verified.   
  In addition, most interpretations focus on the hazard ratio, that is often misinterpreted as the relative risk. 
  In this paper, we introduce an alternative to current methodology for assessing a treatment effect in a two-group situation, not relying on the proportional hazards assumption but assuming proportional risks. Precisely, we propose a new non-parametric model to directly estimate the relative risk of two groups to experience an event under the assumption that the risk ratio is constant over time. In addition to this relative measure, our model allows for calculating the number needed to treat as an absolute measure, providing the possibility of an easy and holistic interpretation of the data. 
  We demonstrate the validity of the approach by means of a simulation study and present an application to data from a large randomized controlled trial  investigating the effect of dapagliflozin on the risk of first hospitalization for heart failure.
		\end{abstract}
		
		%\vskip-.2cm
		
		\parindent 0cm
		
		\maketitle
  
		\vspace{0.5cm}
* Corresponding author: Kathrin M\"ollenhoff, eMail: kathrin.moellenhoff@hhu.de

\section{Introduction} \label{sec0}
%Since at latest the mid 2000s results from randomized controlled trials are considered amongst the highest quality of evidence, being classified as such in 2008 by the GRADE working group \cite{GRADE}. 
%Survival analysis data is the result of many different biomedical studies including randomized controlled trials or cohort studies. 
In medical research, time-to-event data measuring the time until a specific endpoint, e.g. the time until death or the time until occurrence of a particular disease, 
%just to mention a few, 
are very common.
In 2022, nearly 1,000 articles including the phrase ''time-to-event'' were collected on PubMed \cite{Pub23}. Therefore, proper analysis of this type of data is of great interest. Well known models include proportional hazard (PH) models like Cox's PH model \cite{Cox72} or the Weibull PH model \cite{KK12} and proportional odds  (PO) models like the log-logistic model \cite{KK12}, resulting in a focus on the hazard ratio (HR) and the odds ratio (OR) as estimates for the treatment effect. This contradicts the preference of reporting the relative risk (RR) when analysing data given in form of $2\times2$ contingency tables \cite{SG04}. 

The RR is characterized by its easy interpretability: If two groups, say A and B, have a RR of $r$ for an event, then group A is $r$ times more likely to experience the event relative to group B. Consequently, a RR of $1$ means an equal risk for both groups. If $r$ larger than $1$, group A has a larger risk than group B, and vice versa if it is smaller than $1$ \cite{Ros16}.
The RR should always be reported in combination with an absolute risk measure like the numbers needed to treat, to classify the relative measure. Once this has been done, even complex statistical results can be communicated easily \cite{SG04}.

The misinterpretation of the HR as a RR has a long tradition, starting with even basic literature using the two terms interchangeably \cite{KM03,KP02}. However, this is incorrect: both measures indicate the same direction in regards to the treatment effect and hence have similar interpretation, but are technically not the same \cite{SA18}. The HR is a conditional measure, based on rates \cite{Her10}, while the RR is not. 
%Cox's often applied PH model has by definition a time dependent RR for any baseline hazard rate - excluding the trivial case of no treatment effect. The same is true for the Weibull PH model. 
Therefore, the two values should be strictly distinguished and handled with care. 

In addition, an increase in studies reporting non-proportional hazard rates has been noted in recent years \cite{RP14}, putting the PH assumption into question. More precisely, the Cox model and the underlying PH assumption have been criticized recently, as the PH assumption is only rarely assessed in practice, or, the model is even used regardless of the presence of non-proportional hazards \cite{JH19}. 

It is well known that the OR approximates the RR if the event of interest has a low prevalence \cite{SG04}. Otherwise, these measures do not coincide, which led to widespread misinformation in the past \cite{SW99}. Additionally, just as the PH assumption, the PO assumption is sometimes not suitable. 
To fill in the gap of the estimation of the RR for time-to-event data Kuss and Hoyer recently proposed a parametric proportional risk (PPR) model for a two group situation as typically given in in randomized controlled trials by using the exponentiated-uniform distribution \cite{KH21}. This circumvents the problem of the incorrect interpretation of the HR and the OR, respectively. %, both of which are often incorrectly interpreted as RR \cite{SA18}\cite{KD11}.
%\textcolor{blue}{AH: Sind hier eher Probleme bei der richtigen Interpretation des HR und OR gemeint? Das technische Problem einer Annahme (PH, PO oder PR) haben wir ja eigentlich auch...}

Based on this idea, this paper proposes a non-parametric proportional risk (NPPR) estimator for the RR that is easy to implement, robust and particularly independent of the underlying cumulative distribution function. For the construction the Kaplan-Meier estimator \cite{KM58} is used, 
%a well established method, of which the original paper has been cited 61,842 times \cite{Goo23}. This 
 allowing for the inclusion of right censored data.
In addition to a relative risk measure, an absolute measure is needed to enable a holistic interpretation of the data \cite{AcMed17}. Therefore, estimators for the risk difference and the number needed to treat are derived from the NPPR estimator.

The paper is structured as follows: First, we introduce the NPPR model and show how to estimate the treatment effect, given by the mean RR over time. Using this we then present a formula for the risk difference and the number needed to treat. Afterwards, we report a small simulation study that compares the NPPR estimator to the RR estimated by applying the PPR model. Finally, the practical usability of the model  is illustrated by its application to data from the DAPA-HF trial \cite{MS19}, a large randomized controlled trial (RCT) investigating the effect of dapagliflozin on the risk of first hospitalization for heart failure.% in people with type 2 diabetes.

	\section{Methodology} \label{sec1}
\subsection{The non-parametric proportional risk (NPPR) model}
We consider a situation with two groups, in terms of RCTs given by a treatment (indexed by $1$) and a control (indexed by $0$) group, respectively. %The corresponding (unknown) cumulative distribution functions (CDFs) describing the probability of having experienced the outcome event up to a time point over time for each group will be denoted $F_1$ and $F_0$ accordingly. %\textcolor{blue}{CDFs fuer was? Hier sollte spezifiert werden, fuer welche Variable genau eine Verteilung angenommen wird.}
The corresponding (unknown) cumulative distribution functions (CDFs) describing the probability of having experienced the outcome of interest up to a specific time point, are assumed to be proportional. This implies that their ratio, the RR, is constant over time, which we denote as proportional risk (PR) assumption. \\
Let $F_1$ and $F_0$ denote the corresponding unknown CDFs in the treatment and control group, respectively. Given the PR assumption evaluating the ratio of probabilities at any time point $t\geq 0$ (provided $F_0(t)\neq 0$), %\textcolor{red}{OK: Kann ein time point "not well defined" sein? Ich wuerde das weglassen ...} 
yields a treatment effect of
 \begin{align*}
 	\frac{F_1(t)}{F_0(t)}=r
 \end{align*}
 for a constant $r>0$.
 %\textcolor{red}{OK: Die vorherstehende Formel ist eigentlich eine Definition des RR und man muesste das Zeichen := verwenden, oder?}
From this %\textcolor{red}{OK: definition?} 
we define 
%\textcolor{red}{OK: Dann muesste wieder das Zeichen := verwendet werden, oder?} 
 $\beta\in\mathbb{R}$ so that the equation 
  \begin{align*}
 	\exp\left(-\beta\right)=\frac{F_1(t)}{F_0(t)}=r
 \end{align*}
 holds. Solving for $\beta$ we obtain
 \begin{align}\beta&\coloneqq-\log\left(\frac{F_1(t)}{F_0(t)}\right).\label{estsingle}
 	\end{align}
Moving to the log scale allows for a more symmetric interpretation of $\beta$. More precisely, a positive $\beta$ corresponds to a positive treatment effect, a negative $\beta$ to a negative one and $\beta=0$ to no effect, respectively.  The main goal is to estimate $\beta$ without any parametric assumptions on the CDFs.
Therefore, we estimate $F_1$ and $F_0$  %\textcolor{red}{OK: Da bei der Einfuehrung der beiden Gruppen zuerst 1 und dann 0 genannt wurde, wuerde ich die Reihenfolge 1, 0 beibehalten} 
by means of the non-parametric Kaplan-Meier estimators $\hat{S}_1$, $\hat{S}_0$, using the relationship between CDFs and their corresponding survival functions. The resulting estimated CDFs will be denoted by $\hat{F}_i=1-\hat{S}_i$, $i=0,1$. As the underlying data are possibly right censored, we need this slightly more complicated method of estimation instead of using the empirical CDFs. %\textcolor{blue}{AH: Hier muesste der Index i noch eingefuehrt werden.}
\\
Let $T_1$ and $T_0$ be the ordered sets of event time points %(tied observations are excluded here, while still used to estimate $\hat{S}_i$, $i\in\{0,1\}$) \textcolor{blue}{AH: Was sind double values? Sind damit Zeitpunkte gemeint, die mehrfahc auftreten? Die sollten eigentlich nicht geloescht werden...} \textcolor{green}{LA: Ja, genau das ist gemeint. Hier geht es nur darum an wechlen Zeitpunkten wir das Verh\"altnis mit Hilfe der $\hat{F}$ berechnen. Wenn wir das zu einem Zeitpunkt h\"aufiger machen, als zu anderen, haben wir dadurch implizit eine Gewichtung. Im Fall von tied observations sind das genau die Zeitpunkte an denen die CDF gr\"osser springt. Ich sehe nicht, warum genau diese Zeitpunkte einen h\"oheren Informationsgehalt f\"ur das RR haben sollten, als solche mit "einfachen Spr\"ungen". Nat\"Urlichen gehen die tied observations aber noch in die Sch\"atzung des KM also der CDF mit ein und werden daher nicht gel\"oscht. Gerne kann ich sonst auch noch einmal Simulationen mit tied und ohne tied observations testen, erstmal w\"urde ich sie aber heraus lassen.} %) 
for the two groups. %\textcolor{blue}{AH: Waeren das nicht eher observed time points, weil es eventuell keine Events, sondern Zensierungen sind?} \textcolor{green}{Hiermit definieren wir die Menge derZeitpunkte, an denen wir im Schätzer das RR berechnen. Das sollen genau die Sprungstellen der geschätzten CDFs sein, daher hier die Einschränkung auf die event time points.} 
We set $\tilde{t}_{min}=\max(\min(T_1),\min(T_0))$ and $\tilde{t}_{max}=\min(\max(T_1),\max(T_0))$. By defining
 $$\tilde{T}=\left( t\mid t\in T_1\text{ or } t\in T_0, \ \tilde{t}_{min}\leq t\leq \tilde{t}_{max}\right)$$
 we restrict the time interval such that $\beta$ 
%\textcolor{red}{OK: Oben steht, dass das Hauptziel ist, das $\beta$ zu schaetzen und nicht das RR. Natuerlich ist eines eine Funktion des anderen, aber sollte man zum leichteren Verstaendnis vielleicht doch nur eine Bezeichnung nutzen?}
can be evaluated properly. Of note, tied observations are retained in $T_1$, $T_0$ and $\tilde{T}$ for later inclusion. %\textcolor{red}{OK: Diese Bemerkung ueber tied observations taucht etwas ueberraschend auf. Ich denke, man koennte spaeter noch sagen, dass tied observations automatisch mitberuecksichtigt werden, weil diese ja im normalen KM ja auch korrekt beruecksichtigt werden.} \textcolor{green}{LA: Es geht hier nicht um den KM. Wir hatten ja in einem der Gespr\"ache festgehalten, dass wir die tied observations in der Summe, die $\hat{\beta}$ definiert, einzeln aufnehmen wollen. F\"ur n tied observation zum Zeitpunkt taucht t also auch n mal in der Summe auf. Da wir \"uber $\tilde{T}$ summieren, muss das hier erw\"ahnt werden, denn dieser Punkt ist rein Mengentheoretisch etwas unintuitiv.}
%Of note, it might happen that all events of one group are observed before  the first event in the other group occurs. Under those circumstances or when there is no event in one of the groups, $\tilde{T}$ might be empty. However, in this case the PR assumption seems to be not reasonable. 
%We define $\tilde{t}_{min}=\max(\min(T_0),\min(T_1))$ and $\tilde{t}_{max}=\min(\max(T_0),\max(T_1))$. In the following $\tilde{t}_{min}$ will serve as lower bound again in order to avoid dividing by $0$.
%As the continuation of the Kaplan-Meier estimator beyond the time point after the last observed event of the respective group is at discretion of the statistician, we set $\tilde{t}_{max}$ as an upper bound.\\
% If the PR assumption is reasonably supported by the data, \textcolor{red}{OK: Stimmt das? ich denke, es ergeben sich auch KM-Schaetzer ohne "reasonable support" fuer die PR-Annahme. Ich wuerde das streichen und den Satz mit "Inserting ..." beginnen lassen} 
Inserting the estimated CDFs instead of the true unknown CDFs in equation \eqref{estsingle} for any $t$, $\tilde{t}_{min}\leq t\leq \tilde{t}_{max}$, yields a (time-dependent) estimator $\hat{\beta}_t$ for $\beta$. 
 %\textcolor{red}{OK: Das ist aus meiner Sicht nicht korrekt. Ich denke, hier entstehen t verschiedene Schaetzer $\beta_t$ fuer $\beta$. Das $\beta$ an sich bekommen wir erst spaeter aus der Inverse-Variance-Metaanalyse-Anwendung.}
  %\textcolor{blue}{AH: Hier fehlt mir noch der Hinweis, dass jetzt $F_1$ und $F_0$ geschaetzt werden muessen und man das mit dem KM-Schaetzer machen moechte.} \textcolor{green}{LA: Ich habe die Notation entsprechen angepasst und die Reihenfolge etwas ver\"andert.}

\begin{figure}[h]
 	\centering
   \includegraphics[width=.7\textwidth]{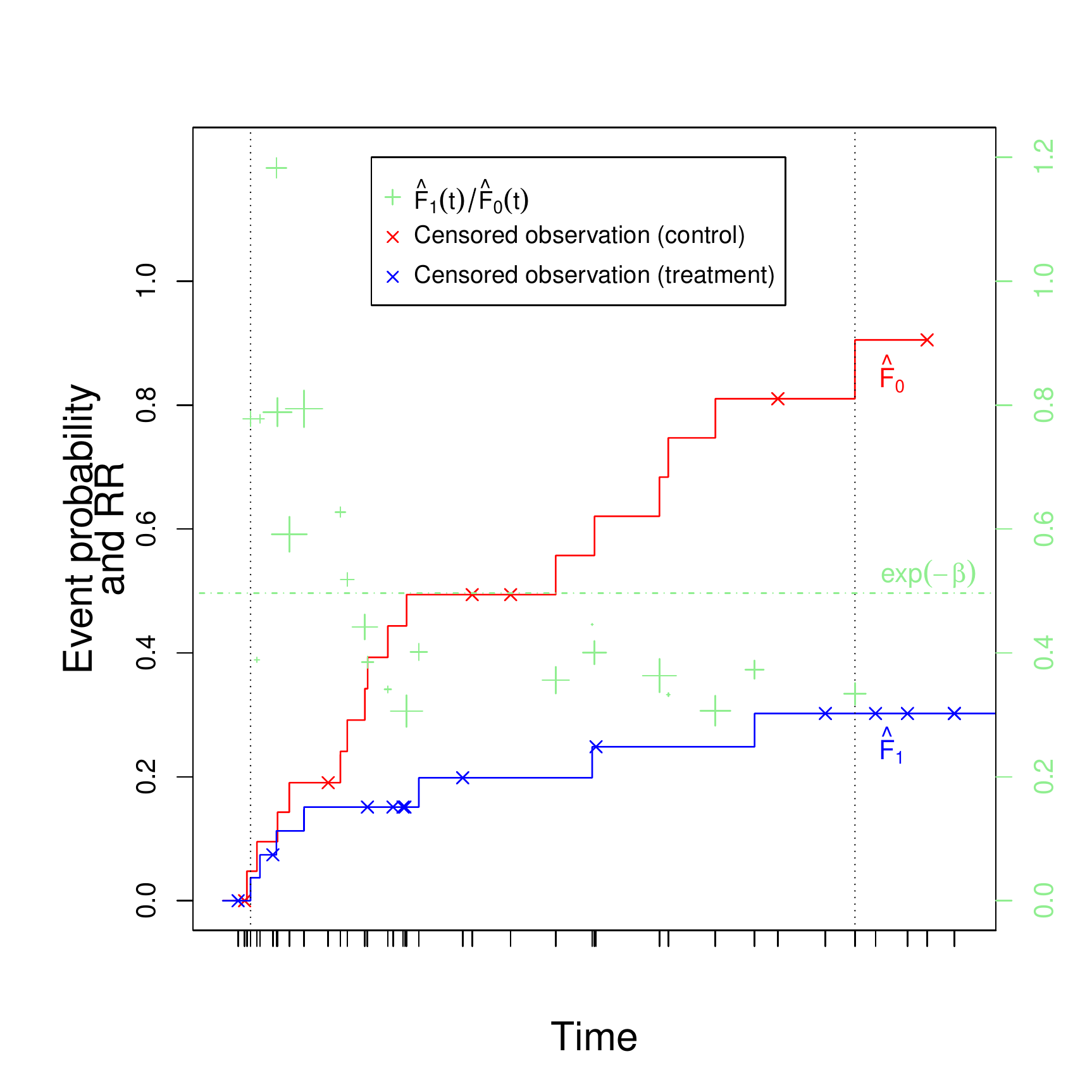}
 	\caption[Figure]{\footnotesize Visualization of the NPPR estimator. The size of the crosses symbolises the weight given to the respective estimated RR at this time point. The larger the cross, the larger the weight.  %The two x-axis describe the contribution  of the treatment ($t`$) and the control ($t$) group to the event points, respectively. 
 % The set of time points was restricted as described hence the first event time point of the treatment group ($1$) being excluded.% At $t7$ we noted $N$ tied observations. The corresponding estimated RR therefore is included $N$ when evaluating the mean. 
  For a visualization on a logarithmic scale we refer to the supplementary material (Figure A).
  }
 	\label{VisEst}
 \end{figure}

 To arrive at a single estimate for $\beta$, we compute a weighted mean of the $\hat{\beta}_t$ where the weights are derived from their variance.
 %However, we need to take into account that the Kaplan-Meier estimator is indeed an estimator and therefore subject to uncertainty. %\textcolor{green}{LA: Wie h\"atten Sie sich an dieser Stelle denn eine Grafik vorgestellt? Ich k\"onnte versuchen eine Art Pfeildiagramm mit den Schritten zu erstellen. Die w\"urde ich dann an das Ende dieses Kapitels setzen, wenn jeder Schritt beschrieben wurde.} 
 %Therefore, we apply this procedure not only at a single time point t, but at multiple %time points, that is every $t\in\tilde{T}$. Afterwards we calculate the weighted mean. %\textcolor{blue}{AH: Koennte man hier noch ein Argument ergaenzen, warum gerade der Mean gewaehlt wird? Kann man das vielleicht mit einfachen Meta-Analysen vergleichen?}
Precisely, as the Kaplan-Meier estimator is asymptotically normally distributed with a variance which can be approximated by Greenwood's formula \cite{BC74}, we estimate the variance of  
 $\hat{\beta}_t=-\log\left(\frac{\hat{F}_1(t)}{\hat{F}_0(t)}\right)=-\log\big(\hat{F}_1(t)\big)+\log\big(\hat{F}_0(t)\big)$ for every $t\in\tilde{T}$ by applying the delta method \cite{Oeh92} to $\log(\hat{F}_i(t))$, $i=0,1$. This yields
 $$\hat{\sigma}^2_i(t)\coloneqq \var\big(\log(\hat{F}_i(t))\big)=\frac{1}{\hat{F}_i(t)^2}\var\big(\hat{S}_i(t)\big), \ \ \ i=0,1.$$
%\textcolor{red}{OK: An manchen Stellen in den Formeln steht $i=0,1$, an anderen werden Mengenklammern verwendet. Das koennte man vielleicht harmonisieren ...}
Since $\hat{F}_1(t)$ and $\hat{F}_0(t)$ are independent and consequently their covariance is equal to $0$, we define the weight function by %$\var\left(-\log\left(\frac{F_1(t)}{F_0(t)}\right)\right)=\var\left(\ln(\hat{F}_1(t))\right)+\var\left(\ln(\hat{F}_o(t))\right)$.
 %\textcolor{blue}{AH: Zur Delta-Methode: Wird die hier wirklich fuer den nicht-parametrischen KM-Schaetzer verwendet? Eigentlich muesste man dafuer die nicht-parametrische Variante verwenden. Koennte man nicht direkt schreiben, dass man die Greenwood-Formel nimmt? Oder steckt hier doch versteckt eine Verteilungsannahme dahinter?} \textcolor{green}{LA: Ich w\"urde an dieser Stelle verwenden, dass der KM Sch\"atzer als ML-Sch\"atzer asymptotisch normalverteilt ist (s.oben)}  To approximate the variance of $\hat{S}_i(t)$, $t\in\tilde{T}$, the Greenwood formula \cite{Gre26} is used.
 %Therefore, the weight function is defined by 
 $$\omega(t)\coloneqq\var\big(\hat{\beta}_t\big)=\frac{1}{\hat{F}_1(t)^2}\var\big(\hat{S}_1(t)\big)+\frac{1}{\hat{F}_0(t)^2}\var\big(\hat{S}_0(t)\big)$$ 
 for  $t\in \tilde{T}$. 
%\textcolor{red}{OK: Sollte man hier noch dazuschreiben, dass die beiden Schaetzer fuer S unabhaengig sind und daher die Kovarianz zwischen beiden wegfaellt?}
 Consequently, with $W:=\sum_{t\in\tilde{T}}\frac{1}{\omega(t)}$, a non-parametric estimator  of $\beta$, the NPPR estimator, is defined by 
 \begin{equation}\label{eq:NPPR}
 \hat{\beta}=\frac{1}{W}\sum_{t\in\tilde{T}}\frac{1}{\omega(t)}\hat{\beta}_t=-\frac{1}{W}\sum_{t\in\tilde{T}}\frac{1}{\omega(t)}\log\left(\frac{\hat{F}_1(t)}{\hat{F}_0(t)}\right).\end{equation}
 As stated above, tied observations were retained in $\tilde{T}$. Therefore, if an event time point $t'$ is included $m$ times over both $T_1$ and $T_0$, the term $\frac{1}{\omega(t')}\hat{\beta}_{t'}$ is also included $m$ times in the sum in \eqref{eq:NPPR}. 
 %While tied observations are also accounted for in the Kaplan-Meier estimator, this is done to include all observations.
 A visualization of the situation is given in Figure \ref{VisEst}. 
% Of note, tied observations are correctly dealt with by virtue of the properties of the Kaplan-Meier estimates. \textcolor{green}{LA: Das ist nicht die Aussage des urspr\"unglichen Satzes hier. Nat\"urlich stimmt es auch, dass tied observations im KM aufgenommen wurden, hier ging es aber darum, dass diese Zeitpunkte dann auch mehrfach in $\tilde{T}$ auftreten und entsprechend mehrfach in der Summe auftreten. }
 %\textcolor{red}{OK: Irgendwo muesste man auch noch einmal schreiben, dass auch Zensierungen adaequat beruecksichtigt werden, oder?}\\
 Also, the estimator does in general estimate the mean RR over time. This holds even if the PR assumption is violated. In this case the estimator still provides a summary of the RR.
% \textcolor{red}{OK: Die beiden vorherstehenden Saetze verstehe ich nicht so ganz. Das geschaetzte $\beta$ ist IMMER ein (gewichtetes) mittleres $\beta$ und damit auch immer ein RR. Das hat aber nichts damit zu tun, ob die PR-Annahme erfuellt ist oder nicht. Anders ausgedrueckt: Wenn die PR-Annahame nicht erfuellt ist, hat man trotzdem einen Schaetzter fuer $\beta$. Ob man mit diesem dann zufrieden ist, wo es doch in Wahrheit kein konstantes $\beta$ gibt, ist etwas anderes. Man koennte aber auch sagen, das gemittelte $\beta$ ist dann trotzdem eine gute Zusammenfassung der $\beta_t$ ...}

 \subsection{Risk difference and number needed to treat}
In order to ensure a holistic interpretation of the data, an absolute measure is needed in addition to the relative measure provided by the RR. The risk difference for the unknown true  CDFs is defined by 
 $$ RD(t)=F_0(t)-F_1(t)$$
 for $t\geq 0$. Using the PR assumption, we can write $F_1(t)=\exp(-\beta)F_0(t)$ and substitute $F_1(t)$ in the formula. This yields
 $$RD(t)=\big(1-\exp(-\beta)\big)F_0(t).$$
 Now we can insert the estimated CDF $\hat{F}_0$ corresponding to the Kaplan-Meier estimator $\hat{S}_0$ and the NPPR estimator $\hat{\beta}$ obtained by \eqref{eq:NPPR}. The resulting estimator of the risk difference is defined by
 $$\widehat{RD}(t)=\big(1-\exp(-\hat{\beta})\big)\hat{F}_0(t)$$
 for $\tilde{t}_{min}\leq t\leq \tilde{t}_{max}$. Of note, this still does depend on the time $t$ and given the PR assumption monotonically decreases if $\exp(-\hat{\beta})<1$ and increases if $\exp(-\hat{\beta})>1$, respectively.
 
 The number needed to treat is defined as the reciprocal of the risk difference. Therefore an estimator is given by
 $$\widehat{NNT}(t)=\frac{1}{\widehat{RD}(t)}=\frac{1}{\big(1-\exp(-\hat{\beta})\big)\hat{F}_0(t)}.$$
% The number needed to harm is the additive inverse of the NNT. Hence an estimator is given by:
 % $$\widehat{NNH}(t)=\frac{1}{\big(\exp(-\hat{\beta})-1\big)\hat{F}_0(t)}.$$
  %The number needed to treat is usually used if a positive treatment effect is expected, while the number needed to harm accounts for negative  effects, e.g. in a toxicological setting. 

 	\section{Simulation study}\label{sec:sim}
   \subsection{Setting and data generation}

% 	\textcolor{green}{LA: Hier ist die Idee unser Modell u.a. gegen die Praxis zu testen, also das Cox's PH model. Genau genommen ben\"otigen wir daher unser Modell auf den Weibull PH Daten (die die PR Annahme auch nicht erfuellen) nicht zwangsl\"aufig, wobei es nat\"urlich sch\"on ist, die Robustheit zu sehen. Die Tabellen sind doch alle etwas sperrig geworden. Vielleicht k\"oenten wir diesen Teil der Tabellen in die Supplements verlegen. Dann sind sie nur noch halb so gross.}
 In order to evaluate the performance of the NPPR estimator compared to an alternative method we conducted a simulation study. 
 %As competitors we chose Cox's PH model, which is most often used in practice, and the PPR model using the exponetiated-uniform (EU) distribution \cite{KH21}. 
 As a competitor we chose the PPR model for estimating the RR. The PPR model is based on the exponentiated-uniform (EU) distribution \cite{KH21}, and the corresponding CDFs are given by
 $$F_{EU,i}(t)=\left(\theta_i t\right)^\alpha, \ \ \ t\leq\frac{1}{\theta_i}, \text{ for } i=0,1,$$ in which the shape parameter $\alpha>0$ is assumed to be the same for both the treatment and the control group and the scale parameters $\theta_i>0$, $i=0,1$, are assumed to be group-specific. By calculating maximum likelihood (ML) estimates using the \texttt{R} function \textit{optim} for $\theta_i$, $i=0,1$, and $\alpha$, the estimated RR is given by
 $$\widehat{RR}_{EU}=\left(\frac{\hat{\theta_1}}{\hat{\theta_0}}\right)^{\hat{\alpha}}.$$ 
 Outcomes of the simulation study were bias, mean squared error (MSE) and empirical coverage of the estimated RR, respectively. Additionally, we evaluated numerical robustness in terms of the number of converged simulation runs. 
 All analyses have been done using R version 4.2.2. 
 %Corresponding R code is available at \textcolor{red}{link}.
 %\textcolor{blue}{AH: Ist es sinnvoll, RR und HR zu vergleichen? Waere es nicht besser, Modelle zu vergleichen, die zum selben Effektmass fuehren? Also z.B. mit dem parametrischen PR-Modell?} \textcolor{green}{LA: Simulationen mit dem parametrischen Modell habe ich erg\"anzt.} 
 
\begin{figure}[h]
 	\centering
 	\includegraphics[width=0.7\textwidth]{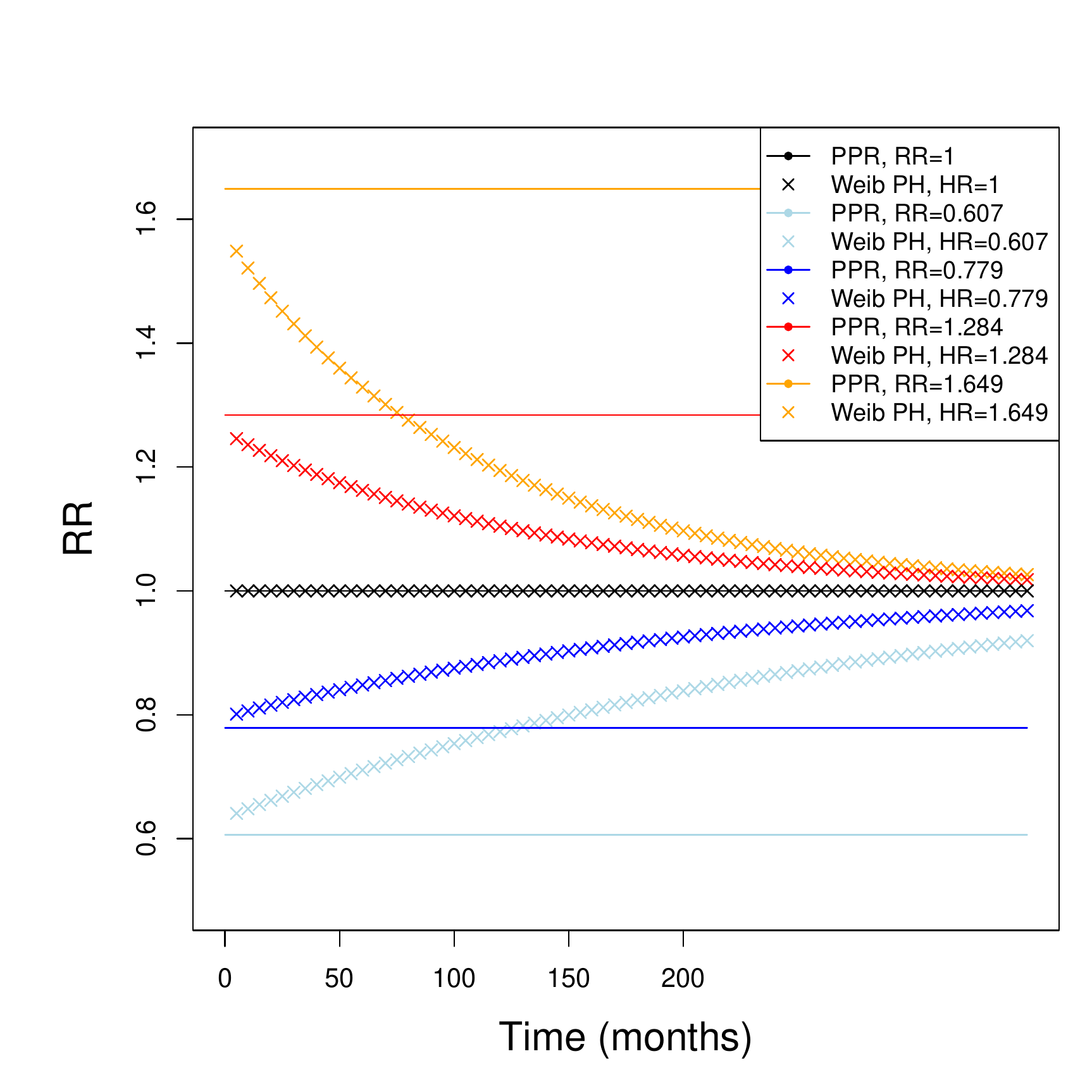}
 	\caption[Figure]{\footnotesize
 	RR of the PPR model in comparison with the RR of the Weibull PH model over time.
 % \textcolor{red}{OK: Im Label wuerde ich statt der Bezeichnung "test" "treatment" verwenden-}
 	%CDFs of the PPR model ($\sim$EU($0.859$, $0.009$))(control), $\sim$EU($0.859$, $0.016$) (treatment, true underlying effect $ 1.649$) and $\sim$EU($0.859$, $0.005$) (treatment, true underlying effect $0.607$) in comparison with the CDFs of the Weibull PH model $\sim$Weib($0.916$, $88.296$) (control), $\sim$Weib($0.916$, $145.575$) (treatment, true underlying effect $ 1.649$) and $\sim$Weib($0.916$, $53.554$) (treatment, true underlying effect $0.607$) with true underlying effect $1$ (left), $ 1.649$ (center) and $0.607$ (right).
 	}
 	\label{rrprph}
 \end{figure}

 The simulation setting was inspired by the DAPA-HF trial discussed as a case study example in Section \ref{sec:casestudy}. We simulated both a PR and a PH situation with different true underlying effects for %$\beta$ 
 % equal to $0$, $0.25$, $0.5$, $-0.25$ and $-0.5$,
%corresponding to a 
the RR %(when simulating data with PR) 
or the HR, %(when simulating data with PH)
respectively. 
%of $1$, $0.779$, $0.607$, $1.284$ and $ 1.649$, respectively. %\textcolor{blue}{AH: Also wird hier angenommen, dass HR und RR identisch sind?} \textcolor{green}{LA: Die Werte waren die gleichen, aber f\"ur NPPR (und EU) f\"ur das RR und f\"ur Weibull PH f\"ur die HR in der Simulation eingesetzt. Finden Sie die Erg\"anzung ausreichend oder ist das zu knapp formuliert? Ansonsten schreibe ich auch diesen Teil noch einmal ausf\"uhrlicher.} 
 For the generation of data satisfying the PR assumption 
 %we took the CDF defining the NPPR model (\ref{nppr}) and made a Weibull $\sim$Weib($0.916$, $88.296$) assumption for the baseline estimated from the placebo group of the data set using the maximum likelihood method. This gave us a parametric model to work with. We will denote it NPPR$_W$. \textcolor{blue}{AH: Das ist mir nicht ganz klar. Wird fuer das Beispiel ein Weibull-PH-Modell geschaetzt und daher kommen die wahren Parameter? DAs NPPR Modell hat doch eigentlich gar keine Weibull-Annahme...}\textcolor{green}{LA: Die Idee war, dass wir f\"ur die Baseline der NPPR Verteilungsfunktion \ref{nppr} eine Weibullannahme getroffen haben. Damit hatten wir dann ein parametrisches Modell, aus dem wir Daten generieren konnten. Ich formuliere diesen Abschnitt noch einmal neu.} Independently 
 the PPR model was used. %The distribution of the control group $\sim$EU($0.859$, $0.009$) was estimated from the placebo group of the data set using the maximum likelihood method and combined with  $\sim$EU($0.859$, $0.007$) for a true underlying effect of $0.779$,  $\sim$EU($0.859$, $0.005$) for a true underlying effect of $0.607$, $\sim$EU($0.859$, $0.012$) for a true underlying effect of $1.284$ and $\sim$EU($0.859$, $0.016$) for a true underlying effect of $ 1.649$.
 For a PH scenario we used a Weibull PH model. The corresponding CDFs are given by
 $$ F_{Weib,i}(t)=\begin{cases} 1-e^{-\left(\frac{t}{\lambda_i}\right)^k} &\text{ for }x\geq 0\\0  &\text{ for }x< 0       \end{cases} \text{ for }i=0,1.$$
 Here, $k>0$ is the shape parameter, which is assumed to be equal for both groups, and $\lambda_{i}$, $i=0,1$, are the group specific scale parameters.
 In both scenarios we used the
 %\textcolor{red}{OK: Was bedeutet "the control group was estimated"? Ich denke, man sollte prinzipiell bei der Beschreibung der Simulation den Teil der Datengenerierung vom Teil der Modellschaetzung trennen. Zum Beispiel wuerde ich auch schreiben, wie (d.h. mit welcher Funktion) das PPR-Modell in R geschaetzt wurde. Wird der R-Code fuer den DAPA-HF trial eigentlich mit publiziert? Faebde ich gut ...} 
   parameters obtained from the placebo group of the data set as underlying truth to simulate data for the reference group,
  % Precisely, using maximum likelihood estimation, we obtained the respective shape parameters of the PPR model ($\alpha$) and the Weibull PH model ($k$) as well as the scale parameters of the control groups ($\theta_{0}$ for the PPR model, $\lambda_{0}$ for the Weibull PH model). 
   %These pairs parameters were then fixed for the models respectively. 
  % The shape parameters of the respective treatment groups ($\theta_{1}$ for the PPR model, $\lambda_{1}$ for the Weibull PH model) were then calculated according to the true underlying effect, 
   where we chose true underlying parameters $\beta=0, \ 0.5,\ 0.25, \ -0.25, \ -0.5 $ which correspond to a RR/HR equal to $1, \ 0.607, \ 0.779, \ 1.284$ and $1.649$ (see Figure \ref{rrprph} for a visualization of the 
  RRs over time).  %and the hazard functions 
 % between the PR and the PH assumption. 
 All configurations used for the simulations are summarized in Table \ref{simparam}. For a visualization of the CDFs we refer to Figure B in the supplementary material. 
   
   %\textcolor{red}{OK: Auch diese Bezeichnung finde ich etwas unscharf. Ich nehme an, dass gemeint ist, dass die Ereigniszeiten in der Behandlungsgruppe simuliert worden sind. Oder bedeutet "constructed" etwas anderes?}
 % with a $\sim$Weib($0.916$, $88.296$) distributed control group, again estimated from the placebo group of the data set using the maximum likelihood method. This is paired with a $\sim$Weib($0.916$, $113.374$) distributed treatment group for a true underlying effect of $ 0.779$, $\sim$Weib($0.916$, $145.575$) for a true underlying effect of $0.607$, $\sim$Weib($0.916$, $68.765$) for a true underlying effect of $1.284$ and $\sim$Weib($0.916$, $53.554$) for a true underlying effect of $1.649$. 
%Details are listed in Table \ref{simparam}. 
To achieve different amounts of censoring ($30\%$, $50\%$ and $70\%$) we simulated censoring times from an appropriate uniform distribution. 
%\textcolor{blue}{AH: Beim vorherigen Satz waere ich etwas genauer: Die Anzahl an zensierten Beobachtungen stammt doch aus einer Bernoulli-Verteilung, oder? Die Gleichverteilung wird dann benutzt, um aus den Event-Zeitpunkten die Zeitpunkte der Zensierung zu machen.} \textcolor{green}{LA: Nein, wir haben Zensurzeitpunkte mit der Gleichverteilung simuliert. Je nachdem ob f\"ur einen Studienteilnehmer dann der Eventzeitpunkt oder der Zensurzeitpunkt fr\"uher auftritt, wurde die Person dann zensiert bzw. nicht.}
For each of these combinations we varied the number of study participants ($50$, $100$ and $500$). All simulation results were obtained from simulating $1,000$ studies.
 Combining both the PR and PH cases resulted in $90$  simulation scenarios representing a situation with no, a realistic and a more extreme effect on both sides of the null effect. 
 In all cases we first drew the group assignment using inverse transformation sampling of a binomial distribution with probability $p=0.5$. Afterwards, corresponding to the declared underlying true effect, 
 we drew survival times from the the PPR model or the Weibull PH model, again using inverse transformation sampling. 
 %\textcolor{red}{OK: Wenn man explizit die Parameter der Weibull-Verteilung hinschreibt, dann sollte man auch die Weibull-Dichte (oder die CDF oder die Hazard-Funktion) einmal hinschreiben. Ansonsten weiss die LeserIn nicht, welche Parametrisierung der Weibull-Verteilung (da gibt es mindestens drei) verwendet wurde.}

 %\textcolor{blue}{AH: Wie geht inverse transform sampling bei einem nicht-parametrischen Modell?} \textcolor{green}{Fuer diesen Schritt hatten wir eine parametrische Annahme fuer die baseline getroffen. Um die Trennung parametrisch/nicht-parametrisch zu verdeutlichen habe ich das abgeaendert und nur mit dem PPR model Daten erzeugt.}
 Given the censoring rate we then drew censoring times from the uniform distribution - for the specific parameters see the supplementary material (Table A) %Table \ref{cens} 
 - for each participant. 
 %\textcolor{red}{OK: Ich habe ehrlicherweise immer noch nicht verstanden, wozu man die Tabelle mit den Gleichverteilungen ueberhaupt braucht. Und selbst wenn, wuerde ich diese ins Supplement packen.}
Finally, the observed time of each individual was defined as minimum of the survival and censoring time and the status adjusted accordingly. 
 %\textcolor{blue}{AH: Insgesamt finde ich setting und data generation noch recht schwierig zu verstehen. Das sollte etwas ausfuehrlicher und genauer beschreiben werden.}

  \begin{table}[h]
 	\footnotesize
 	\centering
 	\caption[Table]{Parameters of the PPR model and the Weibull PH model corresponding to the true underlying effects for all different cases. 
 	}
 	\begin{tabular}{c|c|ccc|ccc}
 		\hline
 		\multicolumn{2}{c|}{True underlying effect} & \multicolumn{6}{c}{Parameters}\\
            \multicolumn{2}{c|}{}&\multicolumn{3}{c|}{PPR}
 		& \multicolumn{3}{c}{Weibull PH}\\
   \hline
   $\beta$&RR/HR& $\alpha$&$\theta_1$&$\theta_0$&$k$&$\lambda_1$&$\lambda_0$\\
 		\hline
0&1&0.859&0.009&0.009&0.916&88.296&88.296\\
0.5&0.607&0.859&0.005&0.009&0.916&145.575&88.296\\
0.25&0.779&0.859&0.007&0.009&0.916&113.374&88.296\\
-0.25&1.284&0.859&0.012&0.009&0.916&68.765&88.296\\
-0.5&1.649&0.859&0.016&0.009&0.916&53.554&88.296
 	\end{tabular}
 	\label{simparam}
 \end{table}
 
 \subsection{Estimation}
 We compared the NPPR estimator with %Cox's PH model and 
 the RR estimated using the PPR model only if for the underlying data generating model the PR assumption was satisfied. For the performance of the PPR model in case of PH data we refer to Kuss \& Hoyer \cite{KH21}. %Analysis on the behavior of Cox's PH model when confronted with PR data was also already conducted there and was not repeated.  % Therefore we fitted all models for the PPR generated data, but in addition to the NPPR model only the Cox's PH model for the Weibull PH generated data. For a comparison of Cox's PH model and the PPR model we refer to the original paper \cite{KH21}. Of note, for the PPR model as true underlying model, the PH assumption is not fulfilled. When fitting the Cox's PH model in this case, we still assumed it to be true, treated the true value of the RR as true value of the HR and proceeded with the analysis accordingly. This was done to represent the use of the Cox's PH model in practice. 
 Of note, the Weibull PH model violates the PR assumption. However, when applying the NPPR estimator we still assumed a PR by treating the true value of the HR as true value of the RR and proceeded with the analysis accordingly. This was done %mimic the use of Cox's PH model in practice and to 
 to examine the behavior of the NPPR model, if the PR assumption does not hold, which can be the case in practical applications. Further, as there is no true underlying constant RR we chose the HR in order to mimic the often seen practice to use the RR and the HR interchangeably \cite{SA18}. 
 
  Of note, it might happen that all events of one group have already occurred before the first event of the other group occurs. Under these circumstances, or when all participants were censored, $\tilde{T}$ might be empty. Then the NPPR estimator is not defined. %\textcolor{blue}{AH: Welche genau?} \textcolor{green}{LA: Ich habe die Beschreibung des Problems aus dem Estimation Kapitel hierher verlegt, sodass es klarer sein sollte.}
 These cases are few in numbers and  were excluded from further analysis. Similarly, for realistic results, datasets yielding estimates obtained by the PPR model of an absolute effect ($|-\log(\widehat{RR}_{EU})|$) larger than $3$ were removed. This choice was informed by an analysis of the simulated data with Cox's PH model, that is not shown here. Precisely, limiting the analysis to estimated HRs smaller than $20$ ($\approx\exp(3)$) was the most conservative way to exclude estimations suffering from obvious numerical issues. Of note, the NPPR estimator exceeded this limit in no case.
 
 Finally, we again refer to the results presented by Kuss \& Hoyer \cite{KH21} for the reversed situation of Cox's PH model used in case of PR data. %For an approximation of the mean RR over time we refer to figure \ref{rrprph}. %\textcolor{red}{Damit ist aber nicht die numerical robustness gemeint, oder? Ich bin mir aber immer noch unsicher, ob es korrekt ist, das wahre HR als wahres RR zu interpretieren. In der Simulation aus dem Weibull-Modell sind ja nicht nur die Annahmen des NPRR-Schaetzers verletzt, sondern man weiss nicht einmal den wahren Wert, der geschaetzt werden soll. Sollte man da noch von Ueberpruefung der Robustheit sprechen?} \textcolor{green}{LA: JA, damit ist nicht numerical robustness gemeint. Hier geht es darum, das oft in der Praxis HR und RR synonym verwendet wird und insbesondere das HR beim Cox model oft als RR interpretiert wird. Daher auch ursprünglich an dieser Stelle der Vergleich zum Cox model. Hier ist Figure 2 auch sehr hilfreich, weil man hier schon sieht, dass es doch das mean RR über die Zeit ist. Die Formulierung habe ich angepasst, damit es klarer ist}
 %The same was done for the NPPR model, when the Weibull PH model was the true underlying model with regards to the PR assumption, the true value of the HR and the estimated RR.
 
 %In order to test the NPPR estimator against the best possible implementation we used the respective true values as starting values for the maximum likelihood estimation of the PPR model.
%As starting values for the maximum likelihood estimation of the PPR model we used the respective true values in order to test the NPPR against the best possible implementation.
 \subsection{Results}
 In the following we will present the results of the simulation study. As the NPPR model directly estimates $\hat{\beta}=-\log(\widehat{RR})$ we transform the estimate $\widehat{RR}_{EU}$ obtained by the PPR model to $-\log(\widehat{RR}_{EU})$ for an easier comparability.
 %An overview of the bias, MSE, coverage and numerical robustness will be given. 
 Table \ref{biasmsepr} and Table \ref{biasmseph} display the bias and the MSE for the different scenarios, respectively. For details on the coverage and the numerical robustness we refer to the supplementary material (Tables B-E).
 %This is the most conservative way to exclude estimations with values higher than thousand - the highest estimate included this way was between $15$ and $16$. 
 %We assumed numerical problems
 %with Cox's PH model 
 %in those cases. %The bound between estimates with high bias and numerical problems was less clear for the PPR model. To assure comparability we again excluded estimates larger than $20$. Of note, no estimate using the NPPR model exceeded this limit. 
 %For details on the number of excluded estimations we refer to the supplementary material on numerical robustness.  %\textcolor{blue}{Warum werden diese Ergebnisse geloescht?}\textcolor{green}{LA: Das HR wurde hier teilweise als eine Zahl in den Tausendern bis hin zu Milliarden gesch\"atzt. Beim MSE sah es in den gleichen Simulationen \"ahnlich aus. Das sind so absurd hohe Werte, dass wir vermuten, dass hier numerischen etwas im Cox Modell nicht gegriffen hat. Diese haben dann die Ergebnisse stark verf\"alscht. Pro Simulationsgruppe sind es immer unter 4 F\"alle. $20$ war schlicht die konservativste Grenze, um diese F\"alle auszuschliessen. Das habe ich entsprechend erg\"anzt.}
 
\begin{figure}[h!]
  	\centering
  	\includegraphics[width=0.8\textwidth]{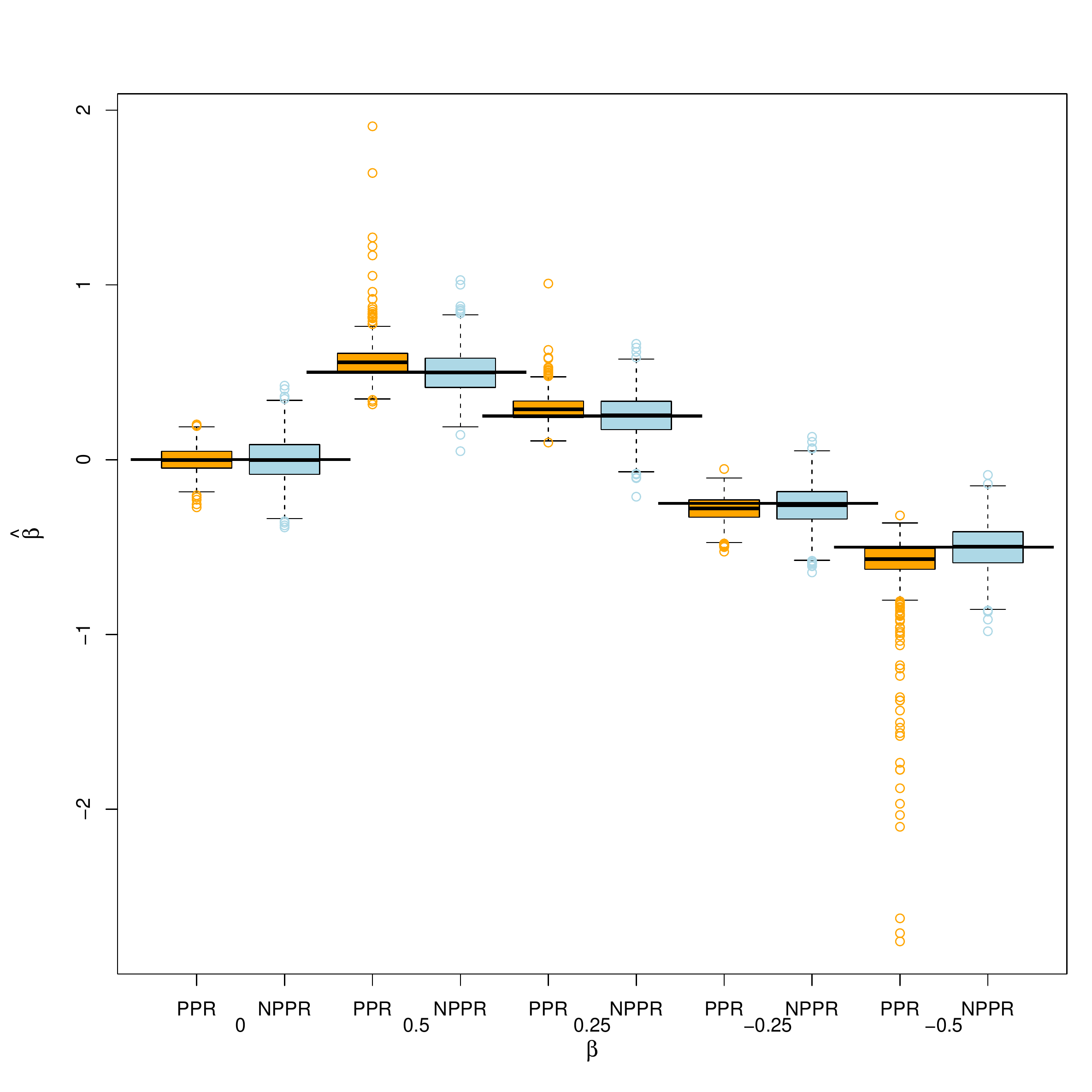}
 	\caption[Figure]{\footnotesize Comparison of boxplots for the estimated $\hat{\beta}$ (NPPR) and $-\log(\widehat{RR}_{EU})$ (PPR) for different choices of $\beta$ if the PPR model is the underlying model, each in case of $500$ participants and $50\%$ censoring rate. Horizontal bars indicate the true underlying effect. Boxplots for $30\%$ and $70\%$ censoring rates and for the PH cases are displayed in the supplementary material (Figures C-G).
 	%\textcolor{blue}{AH: Hier fehlt die Beschriftung der y-Achse.}
  }
 	\label{boxplotpr}
 \end{figure}

 For a quick introduction Figure \ref{boxplotpr} % to \ref{boxplotphlarge}
 displays a comparison between the estimated $\hat{\beta}$ (NPPR) and $-\log(\widehat{RR}_{EU})$ (PPR) for a censoring rate of $50\%$ and $500$ participants if the PR assumption is fulfilled. 
 If the true underlying effect $\beta$ equals $0$, the NPPR estimator has a larger IQR than the PPR model with equal median. In all remaining cases the median is closer to the true value, while the IQR is still a bit larger. We observe that the NPPR estimator produces fewer outliers compared to the PPR model. Similar findings are achieved for situations with a censoring rate of $30\%$ and only in case of a high censoring rate, that is $70\%$, the number of outliers is comparable (see supplementary material Figure C-D). %The performance of the NPPR model is superior for estimation of the RR to Cox's PH model if the PPR model is the true underlying model, as the median is closer to the true value with a comparable IQR and number of outliers. However, the results given by Cox's PH model are consistent with the behavior of the hazard functions (see Figure \ref{prphhazards}). While the PPR model does perform better than the Cox's PH model, the NPPR model still shows better results for a censoring rate of $30\%$ and $50\%$, for any effect not equal to $1$. The PPR model has a comparable median and a smaller IQR in this case for a censoring rate of $50\%$ and $70\%$. For a censoring rate of $30\%$ the NPPR model produces far fewer outliers. Curiously the PPR model does increases in preciseness for higher rates of censoring and performs slightly better than both the NPPR model and Cox's PH model for a censoring rate of $70\%$. In case of the Weibull PH model being the true underlying model, Cox's PH model proves to be superior. Although  the PR assumption is violated (see Figure \ref{rrprph}), it is still a good approximation of the mean RR  over time (see Figure \ref{rrprph}), which demonstrates the robustness of the NPPR model.
 \subsubsection{Bias}

 Tables \ref{biasmsepr} and \ref{biasmseph} depict the bias.  %It becomes obvious that it is important to consider the difference between PR and PH, as both models perform best, when the respective underlying assumption is fulfilled.
 %If both assumptions are fulfilled, the NPPR model performs comparably to Cox's PH model. 
 %The NPPR estimator performs overall satisfactory if the PPR model is the true underlying model. 
 It turns out that if the censoring rate is $50\%$ or smaller the NPPR estimator consistently outperforms the PPR model for all underlying effects but $\beta=0$. 
 This also holds for almost all other configurations, with only a few exceptions. For instance,
 %if $\beta=0.5$,  the bias of the NPPR model is higher for $70\% $ censoring and $50$ participants, for $\beta=-0.5$ only for $100$ participants. 
 considering a censoring rate of $70\%$ and a true effect of $\beta=0.25$ or $\beta=-0.25$ the bias of the NPPR model is for some configurations slightly larger than the one presented by the PPR model. If there is no treatment effect, i.e. $\beta=0$, the difference in biases mostly ranges in order of a magnitude of $0.004$, which is very small. Only in case of a rather high amount of censoring, i.e. $70\%$, and a small sample size of $50$ participants this difference, given by $0.023$, is noticeably larger.  
 %For a true RR of $1.649$ it results in a smaller bias in all cases. For a true effect equal to $1$ it shows better results only for a $30\%$ censoring rate. The bias never surpasses an absolute value of $0.191$ on the RR scale, which occurs for a true effect of $1$ with a $70\%$ censoring rate in case of $50$ participants. This is smaller than the highest bias of $0.991$ observed for the PPR model for a true effect of $1.649$, $30\%$ censoring rate and $100$ participants. The smallest bias overall ($0.001$) on the RR scale is observed for the NPPR estimator for a true effect of $1.284$ in case of $30\%$ censoring and $500$ participants. In regards to the bias observed for $\beta$, it never surpasses an absolute value of $0.045$.
% For smaller rates of censoring the NPPR it outperforms the PPR model. This switches for higher rates of censoring, except for a true underlying effect of $1.649$, where it has a smaller bias in all cases.
% While overall Cox's PH model shows a good performance, especially  for an underlying true effect of $ 1.649$, the bias increases to values larger than $1$ for a censoring rate of $30\%$. Again we note, that the NPPR model does perform better than the PPR model for censoring rates of $30\%$ and $50\%$, but doesn't for $70\%$, for most ture underlying effects. This is most likely due to the PPR model being parametric and therefore overall better equipped to handle higher censoring.

 If the Weibull PH model is the true underlying model, the NPPR estimator has a slightly larger bias when trying to approximate the HR. It still never ecxeeds  $0.177$. The negative values of $\beta$ tend to be overestimated, while the positive ones are underestimated.  %Cox's PH model performs overall better, whereas the NPPR model performs comparably to Cox's PH model in the PR case. However, the bias doesn't exceed an absolute value of $0.237$ in case of a true effect of $ 1.649$. 
This behavior is consistent with the trend observed for the RR over time (see Figure \ref{rrprph}).
We note that the NPPR estimator is more precise for smaller censoring rates and a higher number of participants, as expected. Overall it performs satisfactory and reliable in all cases under consideration.
 %For the NPPR model we conclude that the number of participants and censoring rate have the expected effect on the bias. %The Cox's PH model shows the same behavior for the number of participants, but surprisingly the bias decreases for higher censoring rates if the PPR model is the true underlying model. This results most likely from the violated PH assumption. The same can be observed for the PPR model, possibly due to problems differentiating the estimates with numerical problems and those with high bias and the large number of outliers for a small censoring rate.

 \begin{table}[]
 	\scriptsize
 	\centering
 		\caption[Table]{ Bias and MSE of $\hat{\beta}$ (NPPR) and $-\log(\widehat{RR}_{EU})$ (PPR) if the PPR model is the true underlying model.}
 	\begin{tabular}[H]{c|c|c||c|c||c|c}
 		\hline
 	 Effect & Censoring ($\%$) & Participants & \multicolumn{2}{c}{Bias} &\multicolumn{2}{c}{MSE}\\\hline
   &&&NPPR&PPR&NPPR&PPR\\\hline
0.00&
30&
500&
0.002&
-0.011&
0.010&
0.120\\
0.00&
30&
100&
0.003&
0.018&
0.050&
0.303\\
0.00&
30&
50&
0.010&
-0.006&
0.106&
0.424\\
0.00&
50&
500&
0.000&
0.000&
0.016&
0.005\\
0.00&
50&
100&
-0.005&
0.002&
0.079&
0.037\\
0.00&
50&
50&
0.006&
0.008&
0.156&
0.140\\
0.00&
70&
500&
-0.005&
-0.003&
0.027&
0.016\\
0.00&
70&
100&
0.006&
0.006&
0.135&
0.084\\
0.00&
70&
50&
-0.040&
-0.017&
0.270&
0.187\\
0.50&
30&
500&
0.003&
0.204&
0.013&
0.175\\
0.50&
30&
100&
-0.004&
0.314&
0.063&
0.458\\
0.50&
30&
50&
0.014&
0.384&
0.115&
0.592\\
0.50&
50&
500&
0.001&
0.065&
0.016&
0.016\\
0.50&
50&
100&
0.001&
0.152&
0.079&
0.118\\
0.50&
50&
50&
-0.029&
0.204&
0.171&
0.284\\
0.50&
70&
500&
0.004&
0.006&
0.029&
0.019\\
0.50&
70&
100&
0.006&
0.027&
0.146&
0.091\\
0.50&
70&
50&
-0.043&
0.007&
0.298&
0.265\\
0.25&
30&
500&
-0.006&
0.087&
0.011&
0.130\\
0.25&
30&
100&
0.003&
0.107&
0.052&
0.407\\
0.25&
30&
50&
-0.009&
0.153&
0.106&
0.391\\
0.25&
50&
500&
0.002&
0.041&
0.015&
0.008\\
0.25&
50&
100&
0.002&
0.112&
0.072&
0.091\\
0.25&
50&
50&
0.008&
0.139&
0.159&
0.197\\
0.25&
70&
500&
0.002&
0.001&
0.026&
0.014\\
0.25&
70&
100&
-0.015&
0.000&
0.127&
0.084\\
0.25&
70&
50&
-0.022&
0.012&
0.285&
0.207\\
-0.25&
30&
500&
0.005&
-0.139&
0.011&
0.188\\
-0.25&
30&
100&
-0.001&
-0.136&
0.059&
0.377\\
-0.25&
30&
50&
0.000&
-0.166&
0.106&
0.525\\
-0.25&
50&
500&
-0.013&
-0.032&
0.015&
0.006\\
-0.25&
50&
100&
0.006&
-0.094&
0.076&
0.091\\
-0.25&
50&
50&
0.002&
-0.116&
0.161&
0.162\\
-0.25&
70&
500&
0.002&
0.000&
0.028&
0.017\\
-0.25&
70&
100&
0.006&
0.006&
0.142&
0.091\\
-0.25&
70&
50&
0.038&
-0.003&
0.280&
0.239\\
-0.50&
30&
500&
0.012&
-0.227&
0.011&
0.222\\
-0.50&
30&
100&
-0.010&
-0.323&
0.065&
0.411\\
-0.50&
30&
50&
-0.009&
-0.297&
0.135&
0.552\\
-0.50&
50&
500&
-0.002&
-0.097&
0.017&
0.052\\
-0.50&
50&
100&
0.005&
-0.184&
0.076&
0.145\\
-0.50&
50&
50&
-0.009&
-0.230&
0.169&
0.258\\
-0.50&
70&
500&
-0.003&
-0.006&
0.031&
0.019\\
-0.50&
70&
100&
0.019&
-0.006&
0.157&
0.098\\
-0.50&
70&
50&
0.045&
-0.051&
0.313&
0.240 	

 \end{tabular}
 	\label{biasmsepr}
 \end{table}

 \begin{table}[]
 	\scriptsize
 	\centering
 		\caption[Table]{ Bias and MSE of $\hat{\beta}$ obtained by the NPPR estimator if the Weibull PH model is the true underlying model.}
 	\begin{tabular}[H]{c|c|c||c|c}
 		\hline
 	 Effect & Censoring ($\%$) & Participants & Bias &MSE\\
   \hline

0.00&
30&
500&
0.001&
0.011\\
0.00&
30&
100&
-0.001&
0.052\\
0.00&
30&
50&
-0.002&
0.096\\
0.00&
50&
500&
0.000&
0.015\\
0.00&
50&
100&
-0.009&
0.079\\
0.00&
50&
50&
0.003&
0.163\\
0.00&
70&
500&
0.006&
0.027\\
0.00&
70&
100&
0.015&
0.136\\
0.00&
70&
50&
-0.020&
0.292\\
0.50&
30&
500&
-0.158&
0.036\\
0.50&
30&
100&
-0.156&
0.078\\
0.50&
30&
50&
-0.177&
0.130\\
0.50&
50&
500&
-0.124&
0.032\\
0.50&
50&
100&
-0.122&
0.100\\
0.50&
50&
50&
-0.125&
0.212\\
0.50&
70&
500&
-0.100&
0.040\\
0.50&
70&
100&
-0.113&
0.164\\
0.50&
70&
50&
-0.128&
0.346\\
0.25&
30&
500&
-0.080&
0.017\\
0.25&
30&
100&
-0.096&
0.057\\
0.25&
30&
50&
-0.090&
0.109\\
0.25&
50&
500&
-0.068&
0.018\\
0.25&
50&
100&
-0.072&
0.079\\
0.25&
50&
50&
-0.070&
0.170\\
0.25&
70&
500&
-0.051&
0.031\\
0.25&
70&
100&
-0.055&
0.141\\
0.25&
70&
50&
-0.052&
0.281\\
-0.25&
30&
500&
0.082&
0.016\\
-0.25&
30&
100&
0.071&
0.055\\
-0.25&
30&
50&
0.087&
0.110\\
-0.25&
50&
500&
0.058&
0.018\\
-0.25&
50&
100&
0.052&
0.079\\
-0.25&
50&
50&
0.065&
0.167\\
-0.25&
70&
500&
0.046&
0.030\\
-0.25&
70&
100&
0.085&
0.172\\
-0.25&
70&
50&
0.063&
0.339\\
-0.50&
30&
500&
0.161&
0.037\\
-0.50&
30&
100&
0.175&
0.080\\
-0.50&
30&
50&
0.147&
0.129\\
-0.50&
50&
500&
0.136&
0.032\\
-0.50&
50&
100&
0.133&
0.090\\
-0.50&
50&
50&
0.140&
0.188\\
-0.50&
70&
500&
0.097&
0.040\\
-0.50&
70&
100&
0.106&
0.156\\
-0.50&
70&
50&
0.140&
0.340
 \end{tabular}
 	\label{biasmseph}
 \end{table}

\subsubsection{MSE}
%Again the MSE found in Table \ref{mseprph} is first of all influenced by the underlying true model. Also if both assumptions are fulfilled, the NPPR model again performs slightly better than Cox's PH model overall and for small censoring rates also better than the PPR model if it is the true underlying model. 
Tables \ref{biasmsepr} and \ref{biasmseph} also depict the observed MSE.
If the PPR model was assumed to be the true model we note the same tendencies as seen with the bias. % in regards to a comparison of the two estimators. 
The NPPR estimator presents a smaller MSE for a low censoring rate of $30\%$ in all cases. Considering $50\%$ censoring it is overall smaller or equal to the one observed with the PPR model if $\beta\neq 0$ and for less than $500$ participants if $\beta=0.25$ or $\beta=-0.25$.
The MSE of the NPPR estimator never exceeds $0.313$ (true value $\beta=-0.50$, $70\%$ censoring, $50$ participants). % as expected for small number of participants and a high censoring rate (true value $-0.25$). 
On the other hand, the PPR model shows a MSE up to $ 0.592$ (true value $0.50$, $30\%$ censoring, $50$ participants).  %Cox's PH model can have a MSE up to $3.457$. For the Weibull PH model, the MSE of the NPPR model reaches $1.027$. The highest MSE of higher than $7$ is reached by the PPR model for a true underlying effect of $1.649$. As this coincides with a high number of outliers as seen in Figure \ref{boxplotprlarge}.
%Again the NPPR model performs better than the PPR model for smaller censoring rates and slightly worse for higher censoring ratex except for a true effect og $1.649$, where it peforms better overall.
If the Weibull PH model is the true underlying model, the NPPR estimator presents a similar MSE as for the PR case. It never exceeds $0.346$ (true value $\beta=0.50$, $70\%$ censoring, $50$ participants).
%Also for all models the number of participants have the expected influence. Of note, the Cox's PH models performance gets better for higher censoring rates if the PPR model is the true underlying model as does the PPR model's. We again assume this to be a consequence of the violated PH assumption and numerical reasons, respectively.
Overall we conclude that the MSEs corresponding to the NPPR model show promising behavior, also in case of a violation of the PR assumption.

%\textcolor{red}{OK: Zwei Dinge zum Konfidenzintervall des Schaetzers. Erstens wuerde ich das Konfidenzintervall bereits bei der Einfuehrung der Methode (also in Kapitel 2) vorstellen und nicht erst bei der Simulation. Zweitens bin ich ueberrascht, dass man fuer das KI extra bootstrappen muss, was bei "normalen" AnwenderInnen in der Regel nicht so gut ankommt, da es eher abschreckt. Ich denke, man kann aus der Theorie der Meta-Analyse noch ein geschlossen zu berechnendes Wald-KI hinschreiben. Sollten wir das nicht versuchen?}

\subsubsection{Coverage}%\textcolor{green}{Memo an mich: Einfuegen PPR model}
\label{sec:coverage}
As in general the covariances %$\Cov(\frac{\hat{F}_1(t)}{\hat{F}_0(t)},\frac{\hat{F}_1(t')}{\hat{F}_0(t')})$ 
$\Cov(\hat\beta_t,\hat\beta_{t'})$ 
for $t\neq t'$,  $t,t'\in\tilde{T}$, are unknown and it is therefore impossible to determine the variance of $\hat{\beta}$ in \eqref{eq:NPPR} directly, the confidence intervals for the NPPR estimator were constructed according to the percentile bootstrap approach \cite{Efr81}.
 %\textcolor{green}{LA: Ich halte es nicht f\"ur zielf\"uhrend an dieser Stelle nach einer geschlossenen Form zu suchen. Auch wenn man verwendet, dass der Kaplan-Meier asymptotisch normalverteilt ist, bringt die im Sch\"atzer verwendete Kombination doch grosse Probleme mit sich, wenn man versucht eine solche zu finden. Ich habe einige kurze Versuche in einfachen F\"allen unternommen, die bereits gescheitert sind.} 
Precisely, for each study with $n$ participants we drew an independent sample with replacement of size $n$  from the simulated study. 
%Consequently, each observation was drawn with probability $\frac{1}{n}$. 
The group assignment and status were not changed throughout. From this sample $\hat{\beta}^*$ was estimated again using the NPPR estimator as described in Section \ref{sec1}. This procedure was repeated $500$ times, yielding $\hat{\beta}^{*(1)},\ldots,\hat{\beta}^{*(500)}$. From these values, the empirical $2.5\%$-quantile and $97.5\%$-quantile, respectively, were determined, defining the corresponding $95\%$ confidence interval. For the PPR model we used the multivariate delta method \cite{Vaa98} to construct the confidence intervals.
%\textcolor{red}{Muss das hier genauer beschrieben werden?} \textcolor{blue}{Was genau meinst du hier? Die empirischen Quantile sollten klar sein oder? Oder generell mehr zum Bootstrap?}\\
%The simulated values for the coverage displayed in Table \ref{covprph}  prove the point made regarding the difference between RR and HRs. Cox's PH model in case of a true underlying Weibull PH model proves to be  robust and overall yields good results. On the other hand, if the PPR model is the true underlying model, the coverage decreases to $0.6\%$. The fact that the coverage increases in case of fewer participants most likely results from a wider width of the confidence intervals.

Details on the coverage for each simulated case are presented in the supplementary material (Tables B-C). In general, if the PPR model is the true underlying model, the simulated coverage of the confidence interval for the NPPR estimator is very close to the desired confidence level of $95\%$. It rarely falls below this value,
the smallest value is given by $93.7\%$ and the highest coverage is given by $98.2\%$. The latter is reached in case of a true effect of $-0.25$, $70\%$ censoring and $50$ participants, which underlines the fact that confidence intervals become rather conservative if the number of events is low.

%more precisely the smallest coverage of $93.7$ is observed  for a true effect of $0.25$ in case of $50\%$ censoring and $500$ participants. On the other hand, the highest coverage of $98.2$ is reached in case of a true effect equal to $-0.25$, $70\%$ censoring and $50$ participants, which underlines the fact that confidence intervals become rather conservative if sample sizes are too small or censoring rates too high. 
Overall the PPR model performs comparably. 
%It surpasses $95\%$ a total of $27$ times and reaches $98.5\%$. 
The smallest simulated coverage equals $89.4\%$ and is observed in case of a true effect equal to $0.50$ with a censoring rate  of $30\%$ and $100$ participants.
If the Weibull PH model is the true underlying model the coverage of the HR, approximated by the RR, is overall too low, which is a direct consequence of the violated PR assumption. However, we conclude that also in this case, for most of the configurations the approximation of the $95\%$-level is still rather precise.
%It never falls below $63.9$ which is observed for a true effect of $-0.50$, $30\%$ censoring and $500$ participants. 
%It surpasses $95\%$ $15$ times and reaches $98\%$. This is consistent with the PR assumption being violated.
%The NPPR model shows similar behavior though for the Weibull PH model as true underlying model the coverage never decreases below $60\%$.
%If both assumptions are fulfilled, the models perform equally. The NPPR model and the PPR model perform com5\%arably in all cases without the clear tendencies seen for the Bias and the MSE.
%\textcolor{red}{K: Frage an alle: Sollen wir die Anzahl an Tabellen so lassen oder reduzieren/teilweise grafisch darstellen?}
%\textcolor{blue}{AH: Ich wuerde die grossen Tabellen ins Supplement packen und im Paper selber nur die zentralen Ergebnisse zeigen, z.B. anhand von Boxplots.}
%\textcolor{green}{LA: Ich finde auch die Tabellen sollten in die Supplements oder aufgetrennt werden, damit sie lesbar sind.}

\subsubsection{Numerical Robustness}
 The NPPR estimator proves to be overall robust, independent of the true underlying model. 
 %It never failed more than $16$ of $1000$ times in any case. 
 Assuming the PPR model, failures (less than 16 in 1,000 simulation runs) were only observed in $5$ of the $45$ scenarios. This is also true for the Weibull PH model as underlying model, where the NPPR model only failed a very few times ($<10$), occurring only in case of a high censoring rate of $70\%$ and a small sample size of $50$ participants. Concerning the numerical robustness, the PPR model is clearly outperformed by the NPPR model
 %Any failure at all was observed a total of $34$ in $45$ times and the number of failures went up to $327$ in $1000$.
%The NPPR model and Cox's PH model prove to be overall robust regardless the true underlying model, Cox's PH model slightly more so, than the NPPR model. Neither failed more than $16$ times. For the NPPR model the main influence is found in the number of participants combined with a high censoring rate as was expected. Both models clearly outperform the PPR model in regards to numerical robustness overall. 
, showing a higher robustness in almost every case.
%Only in case of the true effect of $0.607$ with $50$ participants and a censoring rate of $70\%$ is the PPR model slightly more robust than the NPPR estimator.
For the sake of brevity, details are deferred to the supplementary material (Tables D-E).

 	\section{Case study: DAPA-HF trial} \label{sec:casestudy}
To illustrate the NPPR estimator, we use data from the DAPA-HF trial \cite{MS19}. This randomized, double-blind, placebo-controlled trial 
evaluated dapagliflozin, a sodium glucose cotransporter-2 (SGLT-2) inhibitor, for reducing severe cardiovascular outcomes in patients with heart failure and reduced ejection fraction. In 410 sites in 20 countries, 4,744 patients were
treated for a median observation time of 18.2 months. Here we report the results for the trial's primary outcome, time to worsening heart failure or cardiovascular death. Overall, 385 (out of 2,373) patients experienced this outcome in the treatment group, but 500 (out of 2371) in the control group.

As we had no access to the original data, we digitized the Kaplan-Meier estimates from the original paper by the open software tool WebPlotDigitizer, version 3.8 \cite{WP15} and extracted the data by using the algorithms and R tools of Guyot et al. \cite{GA12}. In order to validate this extraction process, we calculated a hazard ratio (with 95\% CI) from the extracted data given by 0.745 [0.653, 0.851], which is essentially the same as the hazard ratio and its CI in the original paper, which was 0.74 [0.65, 0.85].

%\textcolor{red}{OK: Eine Frage zur DAPA-HF-Grafik. Normalerweise werden in KM-Grafiken die zensierten Beobachtungen extra markiert, weil die eigentlichen Events ja durch einen Sprung im KM zu identifizieren sind und daher keine extra Markeirung brauchen. In der Legende steht aber, die Events seien markiert ...}
  
 	\begin{figure}[h!]
 	\begin{center}	\includegraphics[width=.55\textwidth]{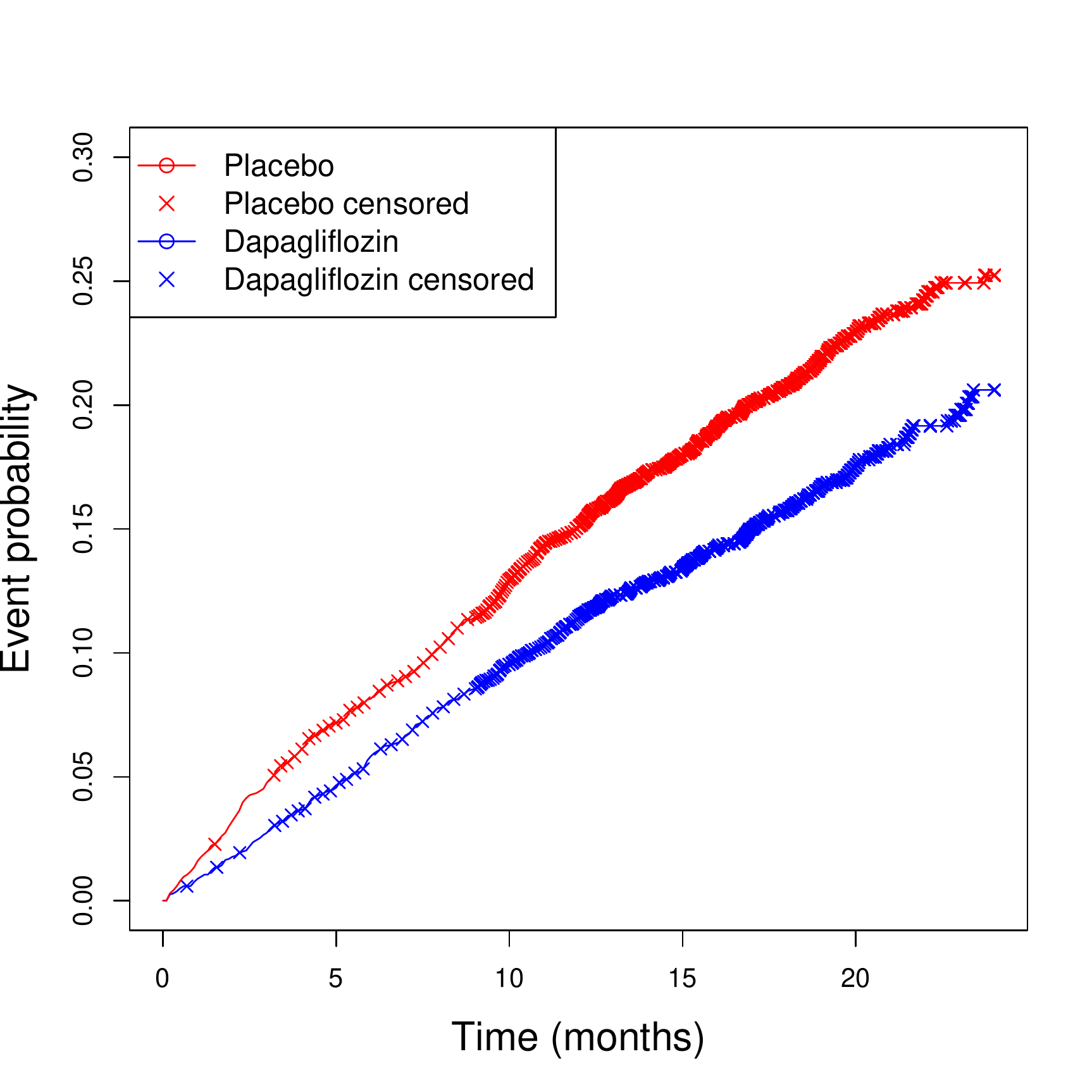}
 		\end{center}
 		\caption[Figure]{\footnotesize  Event probability of the primary outcome of interest and event time points during the DAPA-HF trial estimated with the Kaplan-Meier estimator. }
 		\label{Dapahf}
 	\end{figure}
  
\begin{figure}[h!]
 	\begin{center} \textbf{\footnotesize a)}	\includegraphics[width=.45\textwidth]{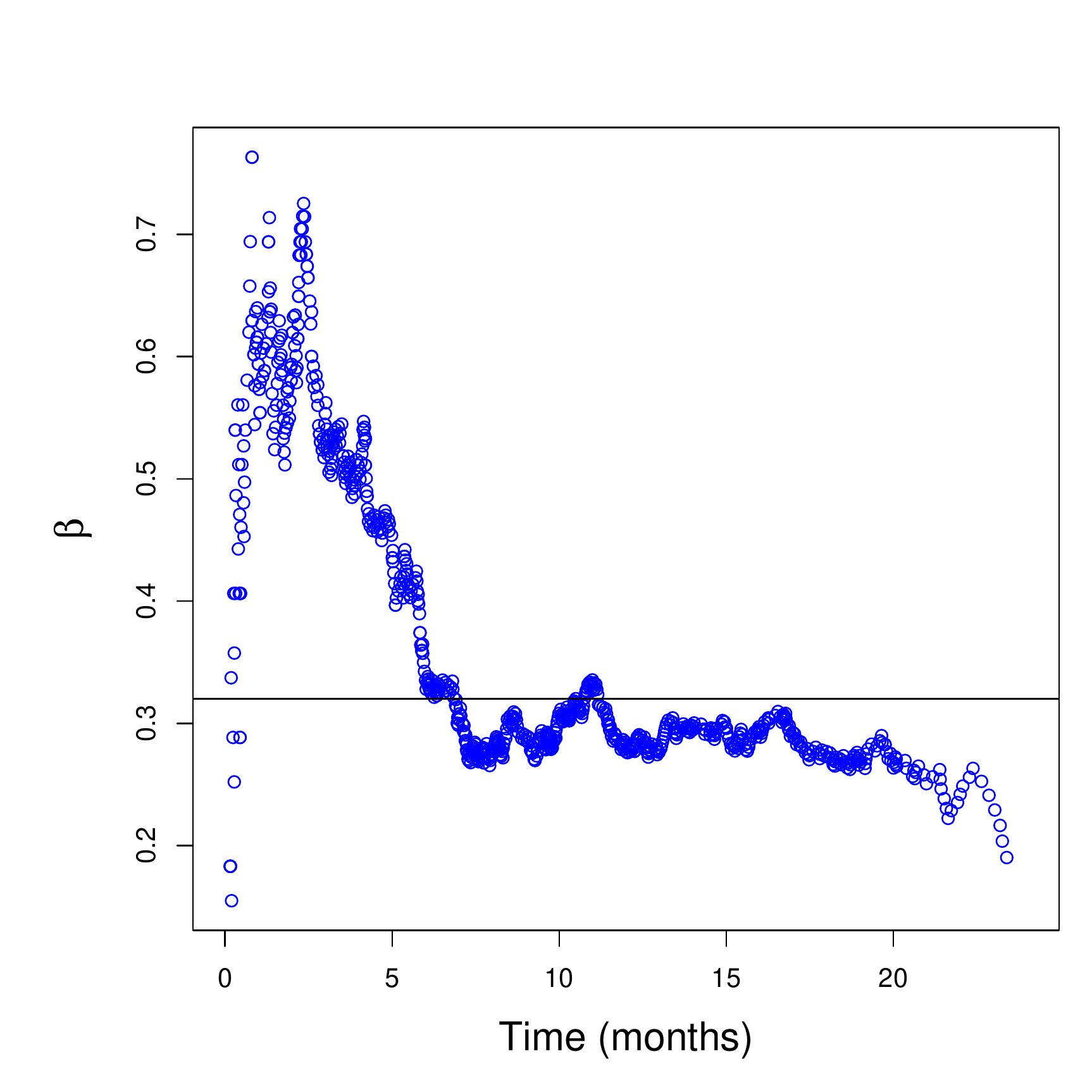}
 	\textbf{\footnotesize b)}	\includegraphics[width=.45\textwidth]{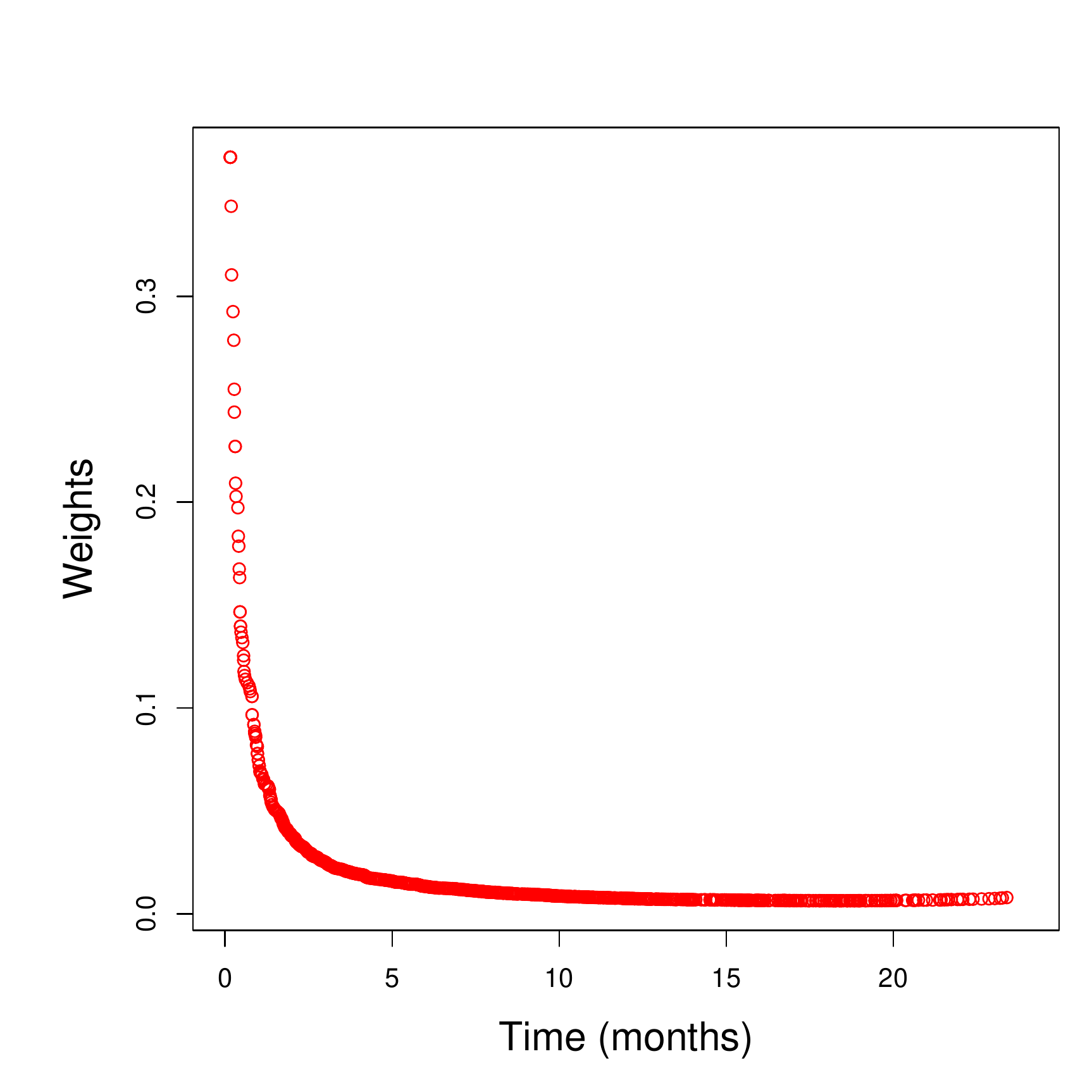}
 		\end{center}
 		\caption[Figure]{\footnotesize \textbf{a)} Estimated $\hat{\beta}_t$  for the DAPA-HF trial at every event time point. The estimated  $\hat{\beta}$ overall is displayed by the line.
 		\textbf{b)} Weights at every event time point. }
   %$\omega(t)=\frac{1}{\hat{F}_1(t)^2}\var\big(\hat{S}_1(t)\big)+\frac{1}{\hat{F}_0(t)^2}\var\big(\hat{S}_0(t)\big)$ } 
 		\label{estdapahf}
 	\end{figure}

  Figure \ref{Dapahf}
 	displays the estimated CDFs from the DAPA-HF trial. The assumption of PR seems to be reasonable. Using the method described in Section \ref{sec1} we obtain $\hat{\beta}= 0.320$ ($95\%$CI $\left[ 	0.178, 0.454\right]$), which corresponds to a RR of $0.726$ ($95\%$CI $\left[ 	0.635,0.837 \right]$). %\textcolor{blue}{AH: 2-3 Nachkommastellen reichen aus.} 
 	The patients in the treatment group are therefore only $72.6\%$ at risk of worsening heart failure or cardiovascular death relative to the patients in the placebo group.
 	 Confidence intervals were estimated using the bootstrap %We assumed random censoring    
%\textcolor{red}{OK: Das kommt etwas ueberraschend. Von censoring bzw. von random censoring war bisher noch gar nicht die Rede gewesen.} \textcolor{green}{LA: Dass die Zensur und die Events unabhängig erfolgen, ist eine Standardannahme. Vermultich ist es daher unnötig es zu erwähnen. Technisch ist es notwendig für den Bootstrap.}
   % and proceeded 
   as described in Section \ref{sec:coverage}. To visualize the estimation process Figure \ref{estdapahf} a) shows the estimated $\hat{\beta}_t$ for each event time point. The weighted mean defining the NPPR estimator as defined in \eqref{eq:NPPR} is displayed by the solid line. The weights at each event time point are shown in Figure \ref{estdapahf} b).
   
%\textcolor{red}{OK: Die beiden Grafiken finde ich super. Allerdings haette ich bei den Gewichten abfallende und nicht ansteigende Werte erwartet. Je spaeter die Zeit, desto kleiner das Sample, desto groesser die Varianzen der KM-Schaetzer, desto kleiner das Gewicht ...}\textcolor{green}{LA: Ich vermute der Effekt entsteht durch den festen Endzeitpunkt bei 24 Monaten. U.a. dadurch haben wir eine sehr hohe Anzahl an zensierten Daten, die erst am "Ende" des Bildes wegfallen. Wir sehen die Varianz erst sinken und dann wieder ansteigen (so wie Sie beschreiben). Am "Rand" nahe der $0$ ist sie nat\"urlich ersteinmal relativ gross.}
    
     The estimated number needed to treat quickly decreases over time, likely due to the small number of reported events in the study.  For instance, at the time point $t=10$ only $29$ ($28.120$ estimated, $95\%$ CI $\left[19.256, 53.947 \right]$) patients treated with dapagliflozin would prevent one death relative to the placebo treatment. A detailed visualization of the number needed to treat are shown in Figure H in the supplementary material.
     
	  %\textcolor{blue}{AH: Mir ist immer noch nicht ganz klar, warum es eigentlich ein nicht-parametrischer Ansatz ist, aber immer mal wieder Verteilungsannahmen gemacht werden... Oder ist es eigentlich ein zweistufiger Prozess? Zuerst wird ueber ML eine parametrische Baseline-Verteilung geschaetzt und ganz unabhaengig davon dann $\beta$ ueber den KM-Schaetzer?} \textcolor{green}{LA: Wir haben hier die Baseline mit einer Verteilungsannahme gesch\"atzt, damit wir unsere Sch\"atzung visualisieren k\"onnen. Alternativ k\"onnte ich das gleiche Ergebnis mit dem Kaplan-Meier fuer die placebo Gruppe erreichen. Dann vergleichen wir aber in \ref{Dapahf} Kaplan-Meier f\"ur die einzelnen Gruppen mit Kaplan-Meier Baseline NPPR. Vermutlich wird das nicht sonderlich sch\"on oder anschaulich aussehen. Als Test kann ich sehr gerne einen Plot erzeugen. Den Punkt mit der Visualisierung habe ich im Text erg\"anzt.} They are compared to the CDFs estimated using the Kaplan-Meier estimator. Both approaches lead to similar curves. This suggests a satisfactory performance of the NPPR model. %However, we note that we cannot draw any conclusions regarding the accuracy of the respective model assumptions, since the underlying truth is not known.

	\section{Discussion and conclusion} \label{concl}
	In this paper we proposed a non-parametric proportional risk model to assess a treatment effect in case of a two-group situation. % and suggested how to extend it to additional binary and categorical variables. 
	 Thereby we solved the conceptional problems the HR and OR have shown in the past without losing the advantages of a non-parametric estimation method. By deducing the number needed to treat from the model, we further provide an absolute measure in addition to the estimated RR. %Indeed an accuracy comparable to the standard Cox's PH model has been demonstrated. 
	As no further specification beyond the PR assumption is necessary, %the model comes without specifying the baseline CDF, 
 the model is broadly applicable and provides a promising tool for the analysis of numerous
	applications, ranging from preclinical toxicology studies to late phase clinical trials, as for instance RCTs. %\textcolor{blue}{AH: Aber in der Simulation und Anwendung braucht man doch immer eine parametrischen Baseline-Verteilung, oder?} \textcolor{green}{LA: Um Daten mit der entsprechenden CDF zu simulieren ja. Ansonsten eher f\"ur \"asthetisch Verwendung wie hier die Visualisierung.}
	
	%\subsection{Limitations and future work}
	Of course, the NPPR estimator comes along with some limitations. Up to this point the model contains only one (binary) variable and it is not possible to include further variables in the model. Moreover the consideration of continuous variables is a challenging problem, which demands some future work. Another drawback, especially if compared to the parametric model, is the necessity of using bootstrap to calculate a confidence interval as a formula for the variance of the estimator is not available yet. 
 %A closed form would proof to be a helpful improvement.\\
%The results are overall satisfactory. 

We demonstrated that the NPPR estimator shows a very good performance if the PR assumption is fulfilled. %and in this case it is overall superior to Cox's PH model. 
For small to moderate censoring rates %with an effect lager tahn and equal to $1$ and for an effect smaller than $1$ overall 
it mostly outperforms the PPR model and is numerically more robust. If the PR assumption is not fulfilled it still shows satisfying behavior.
%consistent with estimating the mean RR over time.
However, if there are only few events, either due to high censoring rates or generally small sample sizes, problems can arise or even make an application impossible. In those cases, while still less robust, the PPR model can provide a better solution.
The NPPR estimator in general estimates the mean RR over time. Therefore, applications outside of a PR scenario might also be of interest.
Further, similarly to the discussion about the PH assumption required from Cox's model, possible problems resulting from a violated PR assumption should be taken into consideration. %One could investigate methods like the introduction of the heaviside function \cite{KK12} for their applicability to the NPPR estimator.%	\textcolor{red}{K: Vielleicht eine Quellenangabe? Oder etwas genauer? } \textcolor{green}{LA: Die Quelle habe ich eingef\"ugt, sollte ich trotzdem einmal die heaviside Funktion definieren und einen Satz zur Anwendung schreiben?}
%\textcolor{red}{OK: Von der heaviside function habe ich ehrlicherweise noch nie gehoert und tue mich daher etwas schwer, das als Erweiterung zu empfehlen. Wenn die PR-Annahme nicht erfuellt ist, dann wuerde ich einfach nicht versuchen, ein RR zu schaetzen ...}
%\textcolor{blue}{Ein anderer Punkt, der mir noch eingefallen ist (aber vielleicht auch gar nicht thematisiert werden muss) ist, dass man das beta ja immer zu diskreten Zeitpunkten berechnet. Es handelt sich also irgendwie um ein Verfahren fuer diskrete Ueberlebenszeiten. In der Praxis ist Stetigkeit aber immer etwas schoener. Waere das nicht eventuell ein Nachteil gegenueber einem parametrischen Modell?}\textcolor{green}{LA: Ich denke, das ist kein großer Punkt, da das NPPR model genau die Event Zeitpunkte verwendet. Die sind unabhängig von jeder Annahme ja zwangsläufig diskret.}

	In summary, we strongly believe that the NPPR estimator is a useful addition to the existing tools for the analysis of time-to-event data,  which not only circumvents the technical problems of the HR and the OR, respectively, but is also easy to interpret and comes along without an assumption on the underlying survival distribution.

 \section*{Supplementary material}

 Supplementary material providing additional tables and figures can be found online. Corresponding R code, which can be used to reproduce the analysis of the case study and the simulation results, is available at \url{https://github.com/LuciaAmeis/NPPR-model-for-time-to-event-data}.

 \section*{Funding}

This work has been supported by the Research Training Group ''Biostatistical Methods for High-Dimensional Data in Toxicology'' (RTG 2624, P7) funded by the Deutsche Forschungsgemeinschaft (DFG, German Research Foundation - Project Number 427806116).
	
%\printbibliography
%\bibliography{LiteraturStat}

\end{document}